\documentclass[ijds,nonblindrev]{informs-ijds}

\OneAndAHalfSpacedXI
\long\def\theARTICLETOPLEFT{{\tabcolsep0pt\fs.12.14.\bf
\begin{tabular}{l}\HD{25}{0}Accepted for publication at the {\bi\theJOURNAL}\\
manuscript {\theMANUSCRIPTNO}\end{tabular}}}%
\RRHSecondLine{\fs.7.9.\rm Article accepted for {\it\theJOURNAL}}% default
\LRHFirstLine{\bf\theRUNAUTHOR:\enskip {\it\theRUNTITLE}}% default
\LRHSecondLine{\fs.7.9.\rm Article accepted for {\it\theJOURNAL}}% default

\def\theARTICLETOP{%
  \vspace*{-57pt}
  \begin{Center}
  \parbox[t][60pt][t]{\textwidth}{%
  \theARTICLETOPLEFT\hfill\theARTICLETOPRIGHT}
  \vspace*{12pt}
  \end{Center}}

\usepackage[no-weekday]{eukdate}

\graphicspath{{figs/}}
\def\addlinespace{}
\usepackage{amsmath,bm,hyperref,url,booktabs}
\usepackage{pdfpages}
\usepackage[section]{placeins}
\providecommand{\tightlist}{%
  \setlength{\itemsep}{0pt}\setlength{\parskip}{0pt}}

\usepackage{natbib}
 \bibpunct[, ]{(}{)}{,}{a}{}{,}%
 \def\bibfont{\small}%
\TheoremsNumberedThrough     % Preferred (Theorem 1, Lemma 1, Theorem 2)
\ECRepeatTheorems

\EquationsNumberedThrough    % Default: (1), (2), ...
\MANUSCRIPTNO{}

\begin{document}

%\setcounter{page}{1}
%\DOI{10.1287/xxxx.0000.0000}%


\RUNTITLE{STR: Seasonal-Trend decomposition using Regression}

% \TITLE{Bundling Information Goods of Decreasing Value}
\TITLE{STR: Seasonal-Trend decomposition using Regression}

\ARTICLEAUTHORS{%
\AUTHOR{Alexander Dokumentov}
\AFF{Let's Forecast, Australia} %, \URL{}}
\AUTHOR{Rob J Hyndman}
\AFF{Department of Econometrics \& Business Statistics, Monash University, Clayton VIC 3800, Australia, \EMAIL{Rob.Hyndman@monash.edu}}
} % end of the block

\ABSTRACT{%
We propose a new method for decomposing seasonal data: STR (a Seasonal-Trend decomposition using Regression). Unlike other decomposition methods, STR allows for multiple seasonal and cyclic components, covariates, seasonal patterns that may have non-integer periods, and seasonality with complex topology. It can be used for time series with any regular time index including hourly, daily, weekly, monthly or quarterly data. It is competitive with existing methods when they exist, but tackles many more decomposition problem than other methods allow.\par
STR is based on a regularized optimization, and so is somewhat related to ridge regression. Because it is based on a statistical model, we can easily compute confidence intervals for components, something that is not possible with most existing decomposition methods (such as STL, X-12-ARIMA, SEATS-TRAMO, etc.).\par
Our model is implemented in the R package \emph{stR}, so can be applied by anyone to their own data.
}%

%\KEYWORDS{deterministic inventory theory; infinite linear programming duality;
%  existence of optimal policies; semi-Markov decision process; cyclic schedule}


\maketitle
%%%%%%%%%%%%%%%%%%%%%%%%%%%%%%%%%%%%%%%%%%%%%%%%%%%%%%%%%%%%%%%%%%%%%%

% Samples of sectioning (and labeling) in MNSC
% NOTE: (1) \section and \subsection do NOT end with a period
%       (2) \subsubsection and lower need end punctuation
%       (3) capitalization is as shown (title style).
%
%\section{Introduction.}\label{intro} %%1.
%\subsection{Duality and the Classical EOQ Problem.}\label{class-EOQ} %% 1.1.
%\subsection{Outline.}\label{outline1} %% 1.2.
%\subsubsection{Cyclic Schedules for the General Deterministic SMDP.}
%  \label{cyclic-schedules} %% 1.2.1
%\section{Problem Description.}\label{problemdescription} %% 2.

% Text of your paper here

\hypertarget{time-series-decomposition}{%
\section{Time series decomposition}\label{time-series-decomposition}}

Decomposition of univariate time series data is a common tool used by government and business in order to understand time series variation. A time series decomposition method separates a time series into components, typically trend, seasonality and remainder. These can be used in many ways for decision-making.

\begin{enumerate}
\def\labelenumi{\arabic{enumi}.}
\tightlist
\item
  It helps separate and understand trend and seasonal effects within a time series. This allows a greater understanding of peak times, and an estimation of the relative effect of holidays, working days and weekends. These can then be used in resource planning.
\item
  It allows the estimation of seasonally-adjusted data, which can then be used to assess the health of a business or industry after adjusting for periodic effects. This is done, for example, with labour market statistics where the seasonally adjusted unemployment rate is used by governments in assessing the state of the economy.
\item
  It can be used to identify anomalies by eliminating the predictable variation due to trend and seasonality, leaving the unpredictable variation (the ``remainder''). Any anomalies in the data are revealed in this ``remainder'' component, when they are often hidden in the original series. Anomaly detection helps organizations identify problems that may need attention. For example, many organizations measure millions of time series at high frequency and some anomaly detection is essential to identify where analysts should address their attention.
\end{enumerate}

Existing time series decomposition methods are designed for monthly and quarterly data, with few tools available for more frequent data. The STL procedure of \citet{cleveland1990stl} is the only widely available decomposition tool for data observed more frequently than monthly, but it assumes a very simple additive structure and allows no covariates. This makes it unsuitable for a large number of problems.

For example, electricity companies measure hourly electricity demand patterns for their customers. Hourly electricity demand data typically shows a time-of-day pattern, a day-of-week pattern and a time-of-year pattern. These seasonalities interact --- the demand pattern on weekdays is different from the demand pattern on weekends, for example. A further complicating issue is that on public holidays, demand tends to look more like weekend patterns than weekday patterns, thus perturbing the periodicity of the patterns. The demand is strongly driven by temperatures (due to the widespread use of electric heating and cooling). Any time series decomposition for electricity demand must allow for all of these effects. However, no existing time series decomposition methods handle this problem, despite this combination of features being extremely common in any human data recorded at a sub-daily frequency.

\hypertarget{a-brief-history-of-time-series-decomposition}{%
\subsection{A brief history of time series decomposition}\label{a-brief-history-of-time-series-decomposition}}

The first attempts to decompose time series into various components can be dated to the 19th century, when \citet{Poynting1884} proposed price averaging as a tool for eliminating trend and seasonal fluctuations. Later, his approach was extended by \citet{hooker1901suspension}, \citet{spencer1904graduation} and \citet{anderson1914nochmals}. Further research in this direction included that by \citet{copeland1915statistical}, who was the first to attempt to extract the seasonal component, until \citet{macaulay1930smooth} proposed a method which gradually became ``classical''. The work of Macaulay led to the Census II method \citep{shiskin1957electronic} which became popular amongst national statistics offices. The Census II method has continued to evolve, with additional features including robustness, calendar effects, covariates, ARIMA extensions, and extensive diagnostics. Widely used versions of this approach have been X-11 \citep{shishkin1967x}, X-11-ARIMA \citep{dagum1988x11arima,ladiray2001seasonal}, X-12-ARIMA \citep{findley1998new} and X-13-ARIMA-SEATS \citep{findley2005some,Dagum2016}.

A different approach was followed by \citet{cleveland1990stl}, who developed STL (Seasonal-Trend decomposition using Loess), which has come to be widely used outside the national statistics agencies, partly because of its availability in R \citep{R}. This method uses iterative Loess smoothing to obtain an estimate of the trend, and then Loess smoothing again to extract a changing additive seasonal component.

Several model-based methods for seasonal decomposition have been developed, including the TRAMO/SEATS procedure which was developed at the Bank of Spain \citep{Gomez2001}, the TBATS model of \citet{delivera2011forecasting}, and various structural time series model approaches \citep{harvey1990forecasting,Commandeur2011}. One big advantage of using a model for seasonal decomposition and adjustment is that it provides a natural way to compute confidence and prediction intervals.

Despite this long history, and the availability of many algorithms and models for time series decomposition, there are many time series characteristics that are not addressed in these approaches. We seek a time series decomposition method with the following attributes:

\begin{itemize}
\tightlist
\item
  provides a meaningful and simple statistical model;
\item
  allows for computation of confidence intervals of components;
\item
  can take account of covariates;
\item
  allows for fractional seasonal periods (e.g., with weekly data);
\item
  allows for multiple seasonal periods (e.g., with daily or sub-daily data);
\item
  allows for complex seasonal topology (e.g., due to public holidays);
\item
  can incorporate covariates which interact with the seasonality.
\end{itemize}

None of the existing methods satisfies all of these requirements. We aim to fill this gap with our new approach, which is clear, general, model-based, robust (if required), and simple. We show that the problem of seasonal decomposition can be re-cast in the framework of ordinary least squares or quantile regression. Moreover, our approach provides new features (such as covariates which affect data in a complex seasonal manner, and the ability to model complex seasonality) that have not been developed before. Our new STR method is the most general framework for the decomposition of seasonal data that is available at present.

\hypertarget{str-models}{%
\section{STR models}\label{str-models}}

We suppose our time series \(y_{t}\) can be decomposed additively as follows:
\begin{equation}
  \label{STR:many_parts}
  y_{t} = T_{t} + \sum_{i=1}^{I} S^{(i)}_{t} + \sum_{p=1}^P \phi_{p,t} z_{t,p} + R_{t} ,
\end{equation}
where

\begin{itemize}
\tightlist
\item
  \(T_{t}\) is a smoothly changing trend;
\item
  \(S^{(i)}_{t}\) are smoothly changing seasonal components with possibly complex topology;
\item
  \(z_{p,t}\) are covariates with coefficients \(\phi_{p,t}\) which may be time-varying and even seasonal;
\item
  \(R_{t}\) is the ``remainder''.
\end{itemize}

The seasonal component is assumed to have a repeating pattern which changes very slowly or is constant over time. The trend component describes the smooth underlying mean of the data (after accounting for seasonality). The remainder component (which we will assume to be uncorrelated) contains only noise and idiosyncratic patterns in the data.

Where components behave multiplicatively, we can use a log or Box-Cox transformation \citep{BC64} to made the additive assumption more reasonable.

The total number of coefficients to be estimated is usually much larger than the number of observations. Consequently, we will impose some regularization on the coefficient estimates. In fact, without some regularization, the trend, seasonal components and time-changing coefficients will be unindentifiable.

The innovation in our approach is to provide an estimation method that allows these regularizations to be imposed in an efficient manner using matrix differencing operators, which convert the estimation method to a linear model. In this way, the estimation becomes analogous to ridge regression. Although the matrices involved can become very large, they are also sparse, so sparse matrix algebra methods can be used to reduce the computational burden.

\hypertarget{smoothness-constraints-via-difference-operators}{%
\subsection{Smoothness constraints via difference operators}\label{smoothness-constraints-via-difference-operators}}

Smoothness of a function implies that its second derivatives are small, since otherwise large values of the second derivatives would lead to high curvature of the function making it ``wiggly'' \citep[see][]{Wood}. Thus, we consider a function smooth if its second derivatives are normally distributed with zero mean and where the variance controls the degree of smoothness.

Because time is discrete, we approximate derivatives with differences. Here, we use the notation \(\Delta_2\) for the double difference operator, \(\Delta_2 = (1-B)^2\), where \(B\) is the backshift operator.

So a smooth trend is obtained by requiring that \(\Delta_2 T_t = T_{t}-2T_{t-1}+T_{t-2}\) are iid \(\mathcal{N}(0,\sigma_L^2)\), where \(\sigma_L\) controls the degree of smoothness. This can be expressed via the multivariate normal density
\[
  f(\bm{D}_\ell \bm{\ell}) = f_L(\sigma_L) \exp\left\{-\frac{1}{2}\big\|\bm{D}_\ell \bm{\ell} / \sigma_L\big\|_{L_2}^2\right\},
\]
where
\(\bm{\ell} = \langle T_{t} \rangle_{t=1}^{n}\) is an \(n\)-vector of the trend component;
\(\bm{D}_\ell\) is the \((n-2) \times n\) difference operator matrix such that \(\bm{D}_\ell\bm{\ell} = \langle\Delta^2 T_{t}\rangle_{t=3}^n\);
and \(f_L(\sigma_L)\) is a normalizing function.

Note that the model subsumes the case of a linear trend, obtained when \(\sigma_L=0\) (making the trend's second derivatives equal to zero).

\hypertarget{smooth-two-dimensional-seasonal-surfaces}{%
\subsection{Smooth two-dimensional seasonal surfaces}\label{smooth-two-dimensional-seasonal-surfaces}}

Let \(m_i\) denote the number of ``seasons'' in the seasonal component \(S^{(i)}_{t}\). For example, if \(y_t\) is monthly data, there is only one seasonal component (\(I=1\)) with \(m_1 = 12\); if \(y_t\) is daily data, there may be two seasonal components (\(I=2\)), with a weekly pattern (\(m_1=7\)) and an annual pattern (\(m_2 = 365\) ignoring leap years). Let us also define the function \(\kappa_i(t) = t\operatorname{mod}m_i\), which transforms time \(t\) into the corresponding season \(\kappa_i(t) \in \{1,\dots,m_i\}\).

At time \(t\), we observe only one element of each seasonal component. It is reasonable to ask what the other elements of the component are at that moment \(t\). For example, if we have daily data with weekly seasonality, it is valid to ask on a Wednesday, ``What is the current value of the seasonal component corresponding to Friday?''. In other words, we define latent components that are responsible for seasons other than \(\kappa_i(t)\). In this way, we treat the \(i\)th seasonal component as a two-dimensional matrix, \([S^{(i)}_{k,t}]\) (\(k=1,\dots,m_i;t=1,\dots,n\)). This formulation of each seasonal component as a two-dimensional structure is one of the novelties and features of STR, and it allows for simple and natural smoothing constraints to be applied.

The seasonal two-dimensional surface actually has the topology of a cylinder, where dimension \(t\) (time) is along the length of the cylinder, but dimension \(s\) (season) is ``circular'' around the cylinder. For example, in the case of a weekly seasonal component in daily data, the season dimension corresponds to days of the week. A time series corresponds to a spiral on the cylinder. The other locations on the cylinder (between the spirals) will represent ``imaginary'' seasonality. These correspond, for example, to the current value of Friday's seasonality when today is a Wednesday.

To ensure they remain identifiable and interpretable, the seasonal terms must sum to zero at any time \(t\), so they must satisfy the property \(\sum\limits_k S^{(i)}_{k,t} = 0\) for each \(t\).

Smoothness of the weekly seasonal pattern can be described as smoothness of \(S^{(i)}_{k,t}\) in the seasonal direction \(k\), while the speed at which the seasonality changes over time is controlled by the smoothness of \(S^{(i)}_{k,t}\) in the time direction \(t\). Moreover, some other features are possible to express. For example, by restricting second partial derivatives of \(S^{(i)}_{k,t}\) in both time and season directions, it is possible to describe the weekly seasonal pattern changing ``synchronously'' in the time-season direction.

The seasonal smoothness can also be expressed using difference operator matrices. We define three such matrices, \(\bm{D}_{tt,i}\), \(\bm{D}_{ss,i}\) and \(\bm{D}_{st,i}\), corresponding to smoothness in the time, season and time-season directions respectively.

For smoothness in the time direction, we require that the seasonal vectors \(\langle \Delta_2 S^{(i)}_{k,t} \rangle_{k=1}^{m_i}\) are iid \(\mathcal{N}(0,\sigma_{i}^2 \bm{\Sigma}_{i} )\), where \(\sigma_i^2\) controls the degree of smoothness and \(\bm{\Sigma}_{i} = \bm{I}_{m_i} - \frac1m\bm{1}_{m_i}\) is the covariance matrix of the \(m_i\) random variables \(\xi_k = \eta_k - m_i^{-1}\sum\limits_{r = 1}^{m_i}\eta_r\), obtained from the iid standard normal random variables \(\eta_1,\dots,\eta_{m_i}\).

Let \(\bm{S}^{(i)} = [S_{k,t}^{(i)}]\) contain the \(i\)th seasonal surface, \(\bm{S}_i^{-}\) be the matrix \(\bm{S}^{(i)}\) without the last row, and let \(\bm{s}_i = \operatorname{vec}(\bm{S}_i^{-})\) be a vector of length \(n(m_i-1)\) representing the \(i\)th seasonal component in vector form. Thus we can write the second differences of the seasonal component along the time dimension as \(\bm{D}_{tt,i} \bm{s}_i = \langle \Delta^2_{t} \bm{S}^{(i)}_{k,t} \rangle_{k=1}^{m_i}\), \(3 \le t \le n\), for a suitably defined \((n-2)m_i \times nm_i\) difference operator matrix \(\bm{D}_{tt,i}\).

Then the density for the second derivatives of the \(i\)th seasonal component is given by
\[
  f(\bm{s}_i)_ =f_{tt,i}(\sigma_i) \exp\left\{-\frac{1}{2}\big\|\ \bm{D}_{tt,i}\bm{s}_i / \sigma_i\big\|_{L_2}^2\right\},
\]
where \(f_{tt,i}(\sigma_i)\) is a normalizing function.

Smoothness in the season and time-season directions are defined analogously by restricting the derivatives \(\frac{\partial^2}{\partial s^2}\) and \(\frac{\partial^2}{\partial s \partial t}\). It should be noted that the matrices \(\bm{D}_{st}\) and \(\bm{D}_{ss}\) take the cylinder topology into account when calculating proper ``circular'' differences.

Again, by setting the smoothing parameters to zero, some simple special cases are obtained. Setting the variance of the seasonal component in the time-seasonal direction to zero causes the seasonal component to be periodic. By setting both trend and time-seasonal variances to zero we effectively force our model to have a linear trend with seasonal dummies. Setting the variance of the second derivatives of the seasonal component in the time direction to zero alone makes changes in the seasonal component linear over time, but does not preclude the seasonal pattern from changing.

\hypertarget{sec:complex-topology}{%
\subsection{STR with complex seasonal topology}\label{sec:complex-topology}}

\begin{figure}[!hbt]
  \centering
  \centerline{Working day \hspace*{4.5cm} Holiday}
  \includegraphics[width=0.6\textwidth]{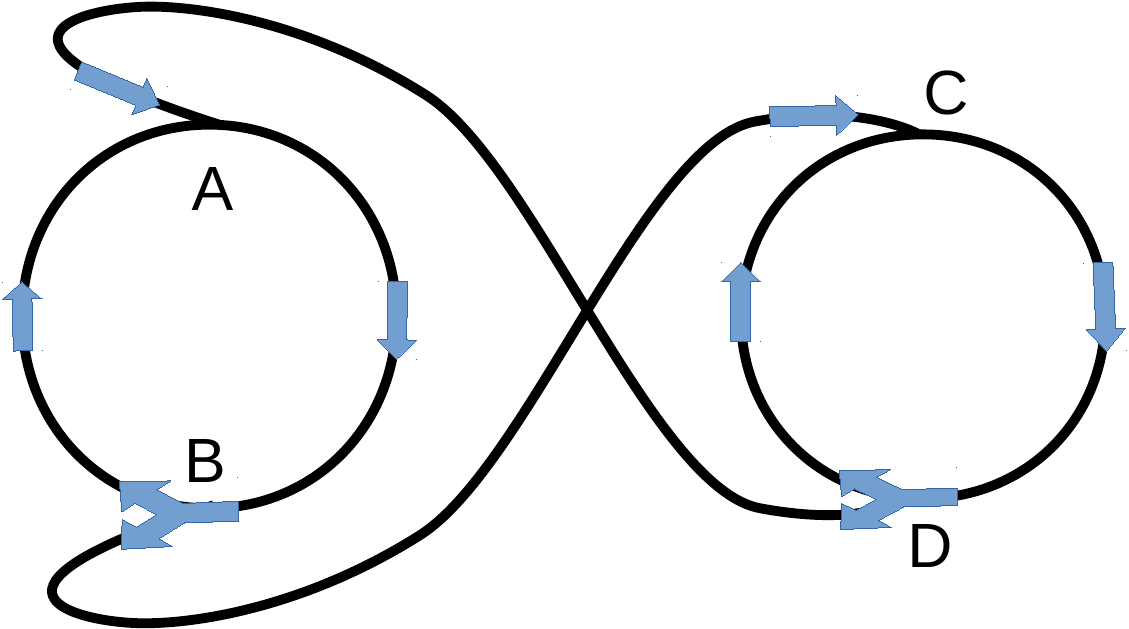}
  \caption{Topology of the complex seasonality modelling human behaviour during working days, holidays and transition periods.}
  \label{fig:topology_pdf}
\end{figure}

A seasonal two-dimensional surface can have a topology other than a cylinder. Suppose that we are going to model some social behaviour (e.g., electricity demand) during working days and holidays (including weekends). The topology modelling the human behaviour is shown in Figure \ref{fig:topology_pdf}.

The left circle represents a working day, the right circle represents a holiday. They are connected by lines representing transition periods. Points A and C represent hour 0 of a day, and points B and D represent hour 12. Every day has 24 hours, and the transition periods take 12 hours.

According to the diagram, a working day can either be followed by another working day, or flow until hour 12 and then undergo a 12 hour transition (line B--C) into a holiday. Similarly, a holiday can either be followed by another holiday, or undergo a 12 hour transition (line D--A) into a working day.

Equivalently, the topology shown in Figure \ref{fig:topology_pdf} can be described as two connected cylinders.

The differencing matrices \(\bm{D}_{ss}\) and \(\bm{D}_{st}\) are defined as above for all data points except A, B, C and D. At points A, B, C and D, the second derivatives can be restricted in various ways. One possibility is to define \(\bm{D}_{ss}\) and \(\bm{D}_{st}\) to regularise the derivatives twice, once for each path in the diagram.

An example of a decomposition with complex topology can be found in Section \ref{sec:one_more_example}.

\hypertarget{gaussian-remainders}{%
\subsection{Gaussian remainders}\label{gaussian-remainders}}

The remainder terms \(R_{t}\) are assumed to be iid \(\mathcal{N}(0,\sigma_R^2)\).
Let \(\bm{y} = [y_1,\dots,y_n]'\) be an \(n\)-vector of observations, \(\bm{Z}=[z_{t,p}]\) be a matrix of covariates with coefficients given by \(\bm{\Phi} = [\phi_{p,t}]\), and let \(\bm{Q}_i\) be an \(n \times n(m_i-1)\) matrix that extracts \(\langle S^{(i)}_{\kappa(t),t} \rangle_{t=1}^{n}\) from the vector \(\bm{s}_i\).

Then the residuals \(\bm{r} = \bm{y} - \sum_i\bm{Q}_i\bm{s}_i -\bm{\ell} - \bm{Z}\bm{\Phi}\) have probability density function
\[
  f(\bm{r}) = f(\sigma_R)\exp\left\{-\frac{1}{2}\big\|\bm{r}/\sigma_R\big\|_{L_2}^2\right\},
\]
where \(f(\sigma_R)\) is a normalizing function.

\hypertarget{covariates-with-time-varying-coefficients}{%
\subsection{Covariates with time-varying coefficients}\label{covariates-with-time-varying-coefficients}}

Covariates are often available to help explain some of the variation seen in a time series. For example, \citet{findley2009stock} considers the effects of various moving holidays on human activities, and \citet{bell2004modeling} considers covariates with time-varying coefficients.

We consider time series that are affected by covariates of three types: (1) static covariates where we assume that the associated coefficients are constant over time; (2) flexible covariates with coefficients whose magnitudes change smoothly over time, but where the coefficients do not exhibit any seasonal pattern; and (3) seasonal covariates with coefficients whose magnitudes change smoothly over time in a seasonal pattern. As far as we know, this last type of time-varying coefficient is new in the decomposition of time series.

To distinguish the three types of covariates, we write
\begin{align*}
  \bm{Z}\bm{\Phi} = \sum_{p=1}^P \phi_{p,t}z_{t,p}
  & =
    \sum_{h=1}^{H} \alpha_{h}{a}_{t,h} +
    \sum_{j=1}^{J} \beta_{j,t}{b}_{t,j} +
    \sum_{k=1}^{K} \gamma_{k,t}{c}_{t,k} \\
  & =
    \bm{A}\bm{\alpha} +
    \sum_{j=1}^{J} \bm{B}_j \bm{\beta}_j +
    \sum_{k=1}^{K} \bm{C}_k \bm{\gamma}_{k},
\end{align*}
where:

\begin{itemize}
\tightlist
\item
  \(a_{t,h}\) are static covariates with constant coefficients;
\item
  \(b_{t,j}\) are flexible covariates with time-varying but non-seasonal coefficients;
\item
  \(c_{t,k}\) are seasonal covariates with time-varying coefficients, where the coefficients have seasonal patterns with corresponding seasonal periods \(p_k \in \{m_1,\dots,m_I\}\);
\item
  \(\bm{A} = [a_{t,h}]\) is an \(n \times H\) matrix of static covariates, where every covariate occupies a single column;
\item
  \(\bm{\alpha}\) is an \(H\)-vector of coefficients of the static covariates;
\item
  \(\bm{B}_j = \operatorname{diag}(\bm{b}_j)\) and \(\bm{b}_j = [b_{1,j},\dots,b_{n,j}]'\);
\item
  \(\bm{\beta}_j\) is the \(j\)th \(n\)-vector of changing coefficients for the \(j\)th flexible covariate;
\item
  \(\bm{C}_k = \operatorname{diag}(\bm{c}_k)\) and \(\bm{c}_j = [c_{1,j},\dots,c_{n,j}]'\);
\item
  \(\bm{\gamma}_k\) is the \(k\)th \(n\)-vector of changing coefficients for the \(k\)th seasonal covariate.
\end{itemize}

To ensure the time-varying coefficients change slowly over time, we will impose smoothness via differencing operators, much as we did for the trend and seasonal components. For this purpose, we define matrices \(\bm{\Delta}_{tt,k}\), \(\bm{\Delta}_{st,k}\) and \(\bm{\Delta}_{ss,k}\) that take second differences of the \(k\)th seasonal coefficients in the time, time-season and season dimensions, respectively.

\hypertarget{maximum-likelihood-estimation}{%
\subsection{Maximum likelihood estimation}\label{maximum-likelihood-estimation}}

\label{sec:STR_solution}

The parameters of the model are given by \(\bm{\alpha}\), \(\bm{\beta}_1,\dots,\bm{\beta}_J\), \(\bm{\gamma}_1,\dots,\bm{\gamma}_K\), \(\bm{\ell}\), \(\bm{s}_1,\dots,\bm{s}_I\), and the standard deviations \(\sigma_R\), \(\sigma_L\), \(\sigma_1,\dots,\sigma_I\) along with those corresponding to the other smoothness constraints.

The model is over-parametrized with more parameters than observations. However, the regularization via the smoothness constraints will allow it to be estimated.

Combining the preceding results, and assuming that all terms are independent of each other and that the standard deviations are known, we find that the minus log likelihood function for this model (up to a constant) is
\begin{align}
  -\log \mathcal{L} &=
  \frac{1}{2\sigma_R}
  \Bigg\{
    \Big\|\bm{y}- \sum_{i=1}^I \bm{Q}_i\bm{s}_i - \bm{\ell} -
    \bm{A}\bm{\alpha} -
    \sum_{j=1}^{J} \bm{B}_j \bm{\beta}_j -
    \sum_{k=1}^{K} \bm{C}_k \bm{\gamma}_{k}
     \Big\|_{L_2}^2
    + \lambda_\ell\Big\|\bm{D}_\ell \bm{\ell}\Big\|_{L_2}^2\nonumber\\
  & \hspace*{1cm}\mbox{} + \sum_{i=1}^{I}\left(
                  \left\|\lambda_{tt,i} \bm{D}_{tt,i} \bm{s}_i \right\|_{L_2}^2 +
                  \left\|\lambda_{st,i} \bm{D}_{st,i} \bm{s}_i \right\|_{L_2}^2 +
                  \left\|\lambda_{ss,i} \bm{D}_{ss,i} \bm{s}_i \right\|_{L_2}^2 \right)
    \nonumber\\
  & \hspace*{1cm}\mbox{} +
           \sum_{j=1}^{J}\left\|\theta_j \bm{D}_{tt} \bm{\beta}_j\right\|_{L_2}^2
    \nonumber\\
  &  \hspace*{1cm}\mbox{}+ \sum_{k=1}^{K}\left(
      \left\|\theta_{tt,k} \bm{\Delta}_{tt,k} \gamma_k\right\|_{L_2}^2
    + \left\|\theta_{st,k} \bm{\Delta}_{st,k} \gamma_k\right\|_{L_2}^2
    + \left\|\theta_{ss,k} \bm{\Delta}_{ss,k} \gamma_k\right\|_{L_2}^2
    \right)
  \Bigg\},
  \label{STR:ll}
\end{align}
where the \(\lambda\) and \(\theta\) coefficients are ratios of standard deviations. Then maximum likelihood estimates can be obtained by minimizing \eqref{STR:ll} over
\begin{equation}\label{regr:beta}
  \bm{\eta} = [\bm{s}_1', \dots, \bm{s}_I', \bm{\ell}', \bm{\alpha}', \bm{\beta}_1', \dots , \bm{\beta}_J', \bm{\gamma}_1', \dots , \bm{\gamma}_K']'.
\end{equation}

Note that \eqref{STR:ll} corresponds to the minus log likelihood function for the linear model
\begin{equation}
  \label{STR:linear_model}
  \bm{y}_{+} = \bm{X}\bm{\eta} + \bm{\varepsilon} ,
\end{equation}
where \(\bm{y}_{+} = [\bm{y}',~ \bm{0}']'\) is the vector \(y\) padded with zeros,
\(\bm{\eta}\) is a vector of unknown coefficients (with seasonal components adjusted by removing the last row of seasonal observations),
\(\bm{\varepsilon} \sim N(\bm{0},\sigma_R^2\bm{I})\),
and
\begin{equation}
  \label{regr:X}
  \bm{X} =
  \begin{bmatrix}
 \bm{Q}_1 & \dots & \bm{Q}_I & \bm{I}_n & \bm{A} & \bm{B}_1 & \dots & \bm{B}_J & \bm{C}_1 & \dots & \bm{C}_K \\
 \lambda_{tt,1} \bm{D}_{tt,1} & \dots & 0 & 0 & 0 & 0 & \dots & 0 & 0 & \dots & 0 \\
 \lambda_{st,1} \bm{D}_{st,1} & \dots & 0 & 0 & 0 & 0 & \dots & 0 & 0 & \dots & 0 \\
 \lambda_{ss,1} \bm{D}_{ss,1} & \dots & 0 & 0 & 0 & 0 & \dots & 0 & 0 & \dots & 0 \\
 0 & \ddots & 0 & 0 & 0 & 0 & \dots & 0 & 0 & \dots & 0 \\
 0 & \dots & \lambda_{tt,I} \bm{D}_{tt,I} & 0 & 0 & 0 & \dots & 0 & 0 & \dots & 0\\
 0 & \dots & \lambda_{st,I} \bm{D}_{st,I} & 0 & 0 & 0 & \dots & 0 & 0 & \dots & 0 \\
 0 & \dots & \lambda_{ss,I} \bm{D}_{ss,I} & 0 & 0 & 0 & \dots & 0 & 0 & \dots & 0 \\
 0 & \dots & 0 & \lambda_\ell \bm{D}_{tt} & 0 & 0 & \dots & 0 & 0 & \dots & 0 \\
 0 & \dots & 0 & 0 & 0 & \theta_1 \bm{D}_{tt} & \dots & 0 & 0 & \dots & 0 \\
 0 & \dots & 0 & 0 & 0 & 0 & \ddots & 0 & 0 & \dots & 0 \\
 0 & \dots & 0 & 0 & 0 & 0 & \dots & \theta_m \bm{D}_{tt} & 0 & \dots & 0 \\
 0 & \dots & 0 & 0 & 0 & 0 & \dots & 0 & \theta_{tt,1} \bm{\Delta}_{tt,1} & \dots & 0 \\
 0 & \dots & 0 & 0 & 0 & 0 & \dots & 0 & \theta_{st,1} \bm{\Delta}_{st,1} & \dots & 0 \\
 0 & \dots & 0 & 0 & 0 & 0 & \dots & 0 & \theta_{ss,1} \bm{\Delta}_{ss,1} & \dots & 0 \\
 0 & \dots & 0 & 0 & 0 & 0 & \dots & 0 & 0 & \ddots & 0 \\
 0 & \dots & 0 & 0 & 0 & 0 & \dots & 0 & 0 & \dots & \theta_{tt,K} \bm{\Delta}_{tt,K} \\
 0 & \dots & 0 & 0 & 0 & 0 & \dots & 0 & 0 & \dots & \theta_{st,K} \bm{\Delta}_{st,K} \\
 0 & \dots & 0 & 0 & 0 & 0 & \dots & 0 & 0 & \dots & \theta_{ss,K} \bm{\Delta}_{ss,K} \\
  \end{bmatrix}
\end{equation}
with fixed parameters
\[
  \bm{\Lambda} = \left\{\lambda_{tt,1}, \lambda_{st,1}, \lambda_{ss,1},
    \dots, \lambda_{tt,r}, \lambda_{st,r}, \lambda_{ss,r}, \lambda_\ell,
    \theta_1, \dots, \theta_J, \theta_{tt,1}, \theta_{st,1}, \theta_{ss,1},
    \dots, \theta_{tt,K}, \theta_{st,K}, \theta_{ss,K}
  \right\}.
\]
If some values of \(\bm{\Lambda}\) are zeros, the corresponding rows of matrix \(\bm{X}\) can (and should) be removed, as they have no effect and removing them improves the computation time.

The total number of coefficients (the length of \(\bm{\eta}\)) that we need to estimate is usually much larger than the number of observations (the length of \(\bm{y}\)). This does not cause computational problems because the coefficients are regularised via smoothness constraints, and the estimation is performed using \eqref{STR:linear_model}, where \(\bm{y}_{+}\) is longer than \(\bm{\eta}\).

Using standard linear regression results, the maximum likelihood solution is given by
\begin{equation}
  \label{STR:solution}
  \hat{\bm\eta} = (\bm{X}'\bm{X})^{-1}\bm{X}'\bm{y}_{+} = (\bm{X}'\bm{X})^{-1}[\bm{Q} ~~ \bm{I}]'\bm{y} .
\end{equation}
with covariance matrix
\begin{equation}
  \label{STR:beta_covar2}
  \text{Cov}(\hat{\bm\eta}) = \sigma_R^2 (\bm{X}'\bm{X})^{-1} .
\end{equation}

The trend component \(\hat{T}_t\) and seasonal components \(\hat{S}_t^{(1)},\dots,\hat{S}_t^{(I)}\), and their corresponding confidence intervals can be obtained directly from \(\hat{\bm\eta}\) and \(\text{Cov}(\hat{\bm\eta})\).

\hypertarget{sec:find_params}{%
\subsection{Smoothing parameter estimation}\label{sec:find_params}}

We will use cross-validation for estimating the \(\lambda\) and \(\theta\) parameters of \eqref{regr:X}. Since the model in \eqref{STR:linear_model} is a linear model, the leave-one-out cross-validation residuals can be calculated \citep[see][]{seber2012linear} using
\begin{equation}
  \label{FindLambdas:residuals}
  \operatorname{e}_{(i)} = \frac{y_i-\hat{y_i}}{1-h_{ii}}\ ,
\end{equation}
where \(y_i\) is the \(i\)th element of the vector \(\bm{y}\), \(\hat{\bm y}_i\) is the \(i\)th element of the vector \(\hat{\bm y} = \bm{H}\bm{y}\), and \(h_{ii}\) is the \(i\)th diagonal element of the hat matrix \(\bm{H} = \bm{X}(\bm{X}'\bm{X})^{-1}\bm{X}'\).
Therefore, we can use the well-known formula for cross-validation in linear regression \citep[see, for example,][p.45]{RWC02}:
\begin{equation}
  \label{FindLambdas:CV}
  \text{CV} = \sum_{i=1}^{n}\left(\frac{y_i-\hat{y_i}}{1-h_{ii}}\right)^2.
\end{equation}

STR finds the optimal \(\lambda\) and \(\theta\) smoothing parameters by minimising CV. The problem of minimising CV can be complex, for example if there are many local minima, and we have no method that guarantees the finding of the global minimum. However, rather simple methods often work well in practice. We perform such optimisation using the Nelder-Mead method as implemented in the \texttt{optim()} function from the stats package in R \citep{R}.

In cases where numerical difficulties may arise in using leave-one-out cross-validation, we resort to \(K\)-fold cross-validation where we split the data set into \(K\) subsets such that the observation at time \(t\) belongs to subset \(i\) if and only if \((t-1) \ \operatorname{mod}\ (K g) \ \in \ [i g, \ \dots \ , (i+1)g-1]\).

When \(g = 1\), no consecutive observations lie in the same subset of data. This gives a reasonable sparsity of the subsets, and we speculate that the resulting \(K\)-fold cross-validation will not differ much from the result of leave-one-out cross-validation. However, in this case, the trend component usually absorbs any serial correlation that may be present in the data.

For \(g>1\), blocks of \(g\) consecutive observations are selected within each subset, similar to a block-bootstrap method, but with a greater separation of neighbouring blocks. The value of \(g\) determines the amount of correlation that is allowed to remain in the residuals, and therefore controls the trade-off between correlated residuals and trend flexibility.

In some cases, we have observed decompositions where the seasonal component ``leaks'' into the trend component. This can be prevented by selecting an appropriate value for \(g\). Setting \(g\) equal to the number of observations in seasonal patterns is an effective choice, as it prevents the trend component from absorbing seasonal patterns with periods less than \(g\).

\hypertarget{simulations}{%
\section{Simulations}\label{simulations}}

To demonstrate that our method estimates the appropriate time series components, we use simulations where we know the true underlying components. We will compare our results against two other decomposition methods that handle multiple seasonal components, STL and TBATS. STL \citep{cleveland1990stl} is a well known decomposition method, and therefore a natural choice to use as a method to compare. The original methodology and implementation was designed for only one seasonal period. However, the implementation in the forecast package for R \citep{Rforecast} will compute an STL decomposition iteratively with multiple seasonal periods. TBATS \citep{delivera2011forecasting} is a forecasting method designed for multiple seasonalities; it provides a decomposition of the time series into components, and so can provide an alternative comparison.

Our simulated daily data consists of four components:
\begin{equation}
  \label{STR:many_parts2}
  y_{t} = T_{t} + \alpha S^{W}_{t} + \beta S^{Y}_{t} + \gamma R_{t} , \quad t=1,\dots,n,
\end{equation}
where \(T_{t}\) is a trend, \(S^{W}_{t}\) is a weekly seasonal component, \(S^{Y}_{t}\) is a yearly seasonal component, \(R_{t}\) is the remainder, and \(\alpha\), \(\beta\), and \(\gamma\) are parameters which control the contribution of the components to \(y_{t}\).

We use two data generating processes (DGPs) to simulate data. For the first DGP, we use a quadratic trend function with random coefficients, such that \(T_t^* = N_1(t + n/2(N_2-1))^2\) where \(N_1\) and \(N_2\) are independent N(0,1) random variables. Then \(T_t^*\) is normalized to give \(T_t\) with mean zero and unit variance. The seasonal components consist of five pairs of Fourier terms with random N(0,1) coefficients, which are then also normalized. Because this DGP has deterministic components, we refer to it as the ``Deterministic'' GDP.

For the second ``Stochastic'' DGP, the trend is given by an ARIMA(0,2,0) model with standard normal errors, which is then normalized. The seasonal components are also smooth and obtained using a similar process constrained to be periodic. Specifically, for each seasonal component we start with a vector of iid N(0,1) values of the length of one season, normalized and replicated to be of length \(n\). This is then integrated and normalized twice to make it smooth. So each seasonal component is as smooth as the trend, is centred on zero and is periodic.

\begin{figure}
\centering
\includegraphics[width=\textwidth]{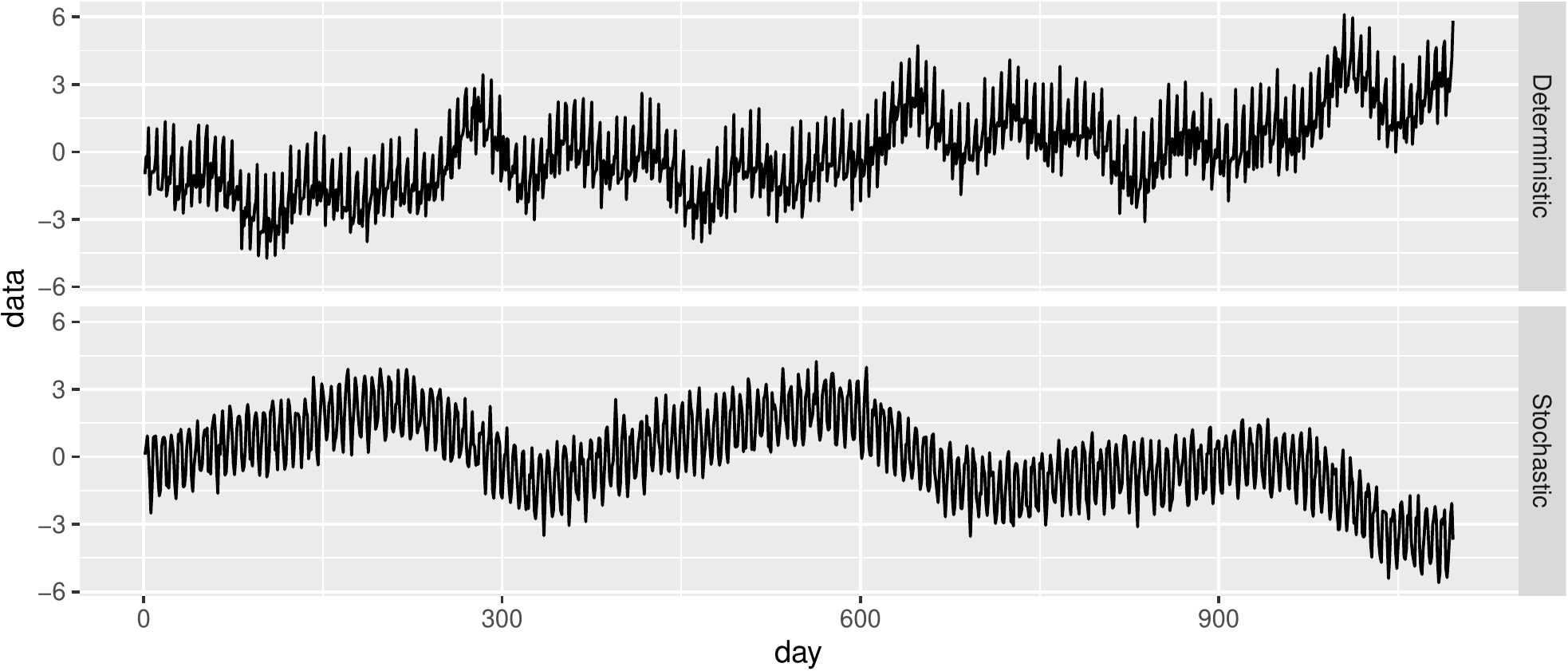}
\caption{\label{fig:examples}Examples of time series generated by the two DGPs.}
\end{figure}

Figure \ref{fig:examples} shows examples of simulated time series from each of the two DGPs with \(\alpha=\beta=1\) and \(\gamma=0.25\). We generated 1096 days in each case (one day more than three non-leap years of daily data).

For comparison of the three decomposition methods, we use parameter sets \((\alpha, \beta, \gamma) = \{(1, 1, 0.2), (1, 1, 0.4), (1, 1, 0.6)\}\), and create 20 time series for each DGP and parameter set. We apply STR, STL and TBATS to each series and compute RMSE for each component (averaging across all data sets and days). We measure the significance of the differences between the methods using a linear model applied to the squared errors. The results are shown in Table \ref{tab:cbetable}. It is clear that STR significantly outperforms STL and TBATS in estimating all components.

\begin{table}[!ht]

\caption{\label{tab:cbetable}RMSE over 100 simulations for each DGP and value of $\gamma$. Bold values indicate results that are significantly different from the STR values.}
\centering
\begin{tabular}[t]{rlrrrr}
\toprule
$\gamma$ & Method & Trend RMSE & Weekly RMSE & Yearly RMSE & Remainder RMSE\\
\midrule
\addlinespace[0.3em]
\multicolumn{6}{l}{\textbf{Stochastic DGP}}\\
\hspace{1em}0.2 & STL & \textbf{0.0808} & \textbf{0.0564} & \textbf{0.1267} & \textbf{0.1525}\\
\hspace{1em}0.2 & TBATS & \textbf{0.1830} & \textbf{0.0174} & \textbf{0.1767} & \textbf{0.0566}\\
\hspace{1em}0.2 & STR & 0.0183 & 0.0149 & 0.0475 & 0.0515\\
\midrule
\hspace{1em}0.4 & STL & \textbf{0.0798} & \textbf{0.1148} & \textbf{0.2299} & \textbf{0.2659}\\
\hspace{1em}0.4 & TBATS & \textbf{0.3435} & \textbf{0.0344} & \textbf{0.3310} & \textbf{0.1227}\\
\hspace{1em}0.4 & STR & 0.0320 & 0.0301 & 0.0778 & 0.0876\\
\midrule
\hspace{1em}0.6 & STL & \textbf{0.0846} & \textbf{0.1717} & \textbf{0.3388} & \textbf{0.3884}\\
\hspace{1em}0.6 & TBATS & \textbf{0.5107} & \textbf{0.0508} & \textbf{0.4983} & \textbf{0.1648}\\
\hspace{1em}0.6 & STR & 0.0523 & 0.0438 & 0.1136 & 0.1273\\
\midrule
\addlinespace[0.3em]
\multicolumn{6}{l}{\textbf{Deterministic GDP}}\\
\hspace{1em}0.2 & STL & \textbf{0.1272} & \textbf{0.0575} & \textbf{0.1402} & \textbf{0.1748}\\
\hspace{1em}0.2 & TBATS & \textbf{0.2304} & \textbf{0.0166} & \textbf{0.2259} & \textbf{0.0803}\\
\hspace{1em}0.2 & STR & 0.0851 & 0.0147 & 0.1000 & 0.0607\\
\midrule
\hspace{1em}0.4 & STL & \textbf{0.1135} & \textbf{0.1147} & \textbf{0.2374} & \textbf{0.2757}\\
\hspace{1em}0.4 & TBATS & \textbf{0.3393} & \textbf{0.0340} & \textbf{0.3294} & \textbf{0.1153}\\
\hspace{1em}0.4 & STR & 0.0584 & 0.0306 & 0.0937 & 0.0929\\
\midrule
\hspace{1em}0.6 & STL & \textbf{0.1439} & \textbf{0.1714} & \textbf{0.3427} & \textbf{0.4018}\\
\hspace{1em}0.6 & TBATS & \textbf{0.5444} & \textbf{0.0509} & \textbf{0.5298} & \textbf{0.1709}\\
\hspace{1em}0.6 & STR & 0.0757 & 0.0436 & 0.1691 & 0.1709\\
\bottomrule
\end{tabular}
\end{table}

\hypertarget{applications}{%
\section{Applications}\label{applications}}

We will illustrate our proposed method using two data sets. The first is a simple application with monthly data, that allows us to compare our results with existing methods. The second is more complicated and demonstrates the full range of capabilities of our method including multiple seasonal patterns, complex seasonal topology due to public holidays and interacting seasonalities, with a nonlinear covariate.

\hypertarget{ssec:simple}{%
\subsection{Monthly supermarket revenue}\label{ssec:simple}}

\begin{figure}
\centering
\includegraphics[width=\textwidth]{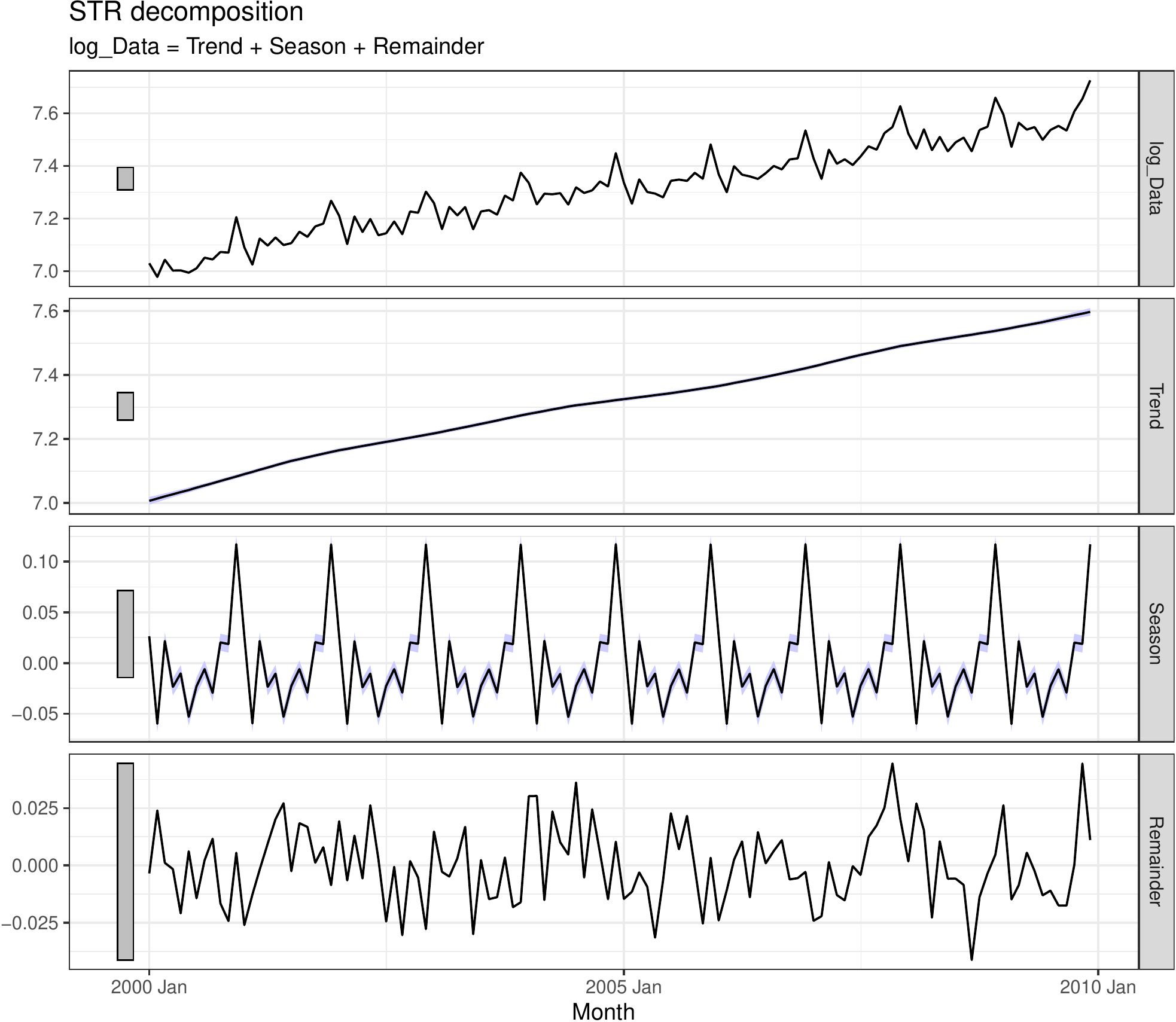}
\caption{\label{fig:supermarket}Supermarket and grocery store log revenue in New South Wales, decomposed with STR. The original data (logged) are shown in the top panel, while the components are shown in the other panels, with 95\% confidence intervals for the trend and season components in blue}
\end{figure}

The time series shown in the top panel of Figure \ref{fig:supermarket} concerns logarithms of monthly supermarket and grocery store revenue in New South Wales, Australia, from 2000 to 2009.

We apply the model in \eqref{STR:linear_model} to the logged data with no covariates and one seasonal component (\(m_1=12\)), with the smoothing parameters selected using leave-one-out cross-validation. The resulting decomposition is shown in Figure \ref{fig:supermarket}, along with 95\% confidence intervals shown as blue bands. Here, the trend component measures the overall health and size of the economy; the seasonal component allows the study of how human behaviour changes over time; the remainder component allows us to identify unusual features of the data.

We obtained similar results using other decomposition methods applied to the same data including STL \citep{cleveland1990stl}, TBATS \citep{delivera2011forecasting} and X-13-ARIMA-SEATS \citep{findley2005some,Dagum2016}. These comparisons are provided in the appendix.

\FloatBarrier

\hypertarget{sec:one_more_example}{%
\subsection{Half-hourly electricity consumption with temperature predictors}\label{sec:one_more_example}}

We illustrate the power of the STR approach by presenting a more complicated example of time series decomposition using half-hourly electricity consumption in the state of Victoria, Australia, during the 115 days starting on 10 January 2000.

For each 30-minute period, we also have the air temperature at the Melbourne weather station (near the centre of the largest city in Victoria). We use concurrent temperatures and their squared values as predictors. Because the effect of temperature can change over time (a mild day in summer will have a different effect from a day of the same temperature in winter), we allow for changing coefficients over time.

This data set has two seasonal patterns. The first pattern is a weekly seasonal pattern (\(m_1=48\times7=336\)) that represents the specific demand features that are attributable to a particular day of the week. The second pattern is a daily seasonal pattern (\(m_2=48\)) with the topology in Figure \ref{fig:topology_pdf}, which allows the model to distinguish between working days and holidays/weekends, and to make transitions between them. The pattern reflects the tendency to have a higher electricity demand during standard working hours and a lower demand at night. It also reflects the tendency to have different demand patterns on working days and holidays/weekends. A longer series would also have an annual seasonal pattern, but with only 115 days, this cannot be distinguished from the trend component.

\begin{figure}
\centering
\includegraphics[width=\textwidth]{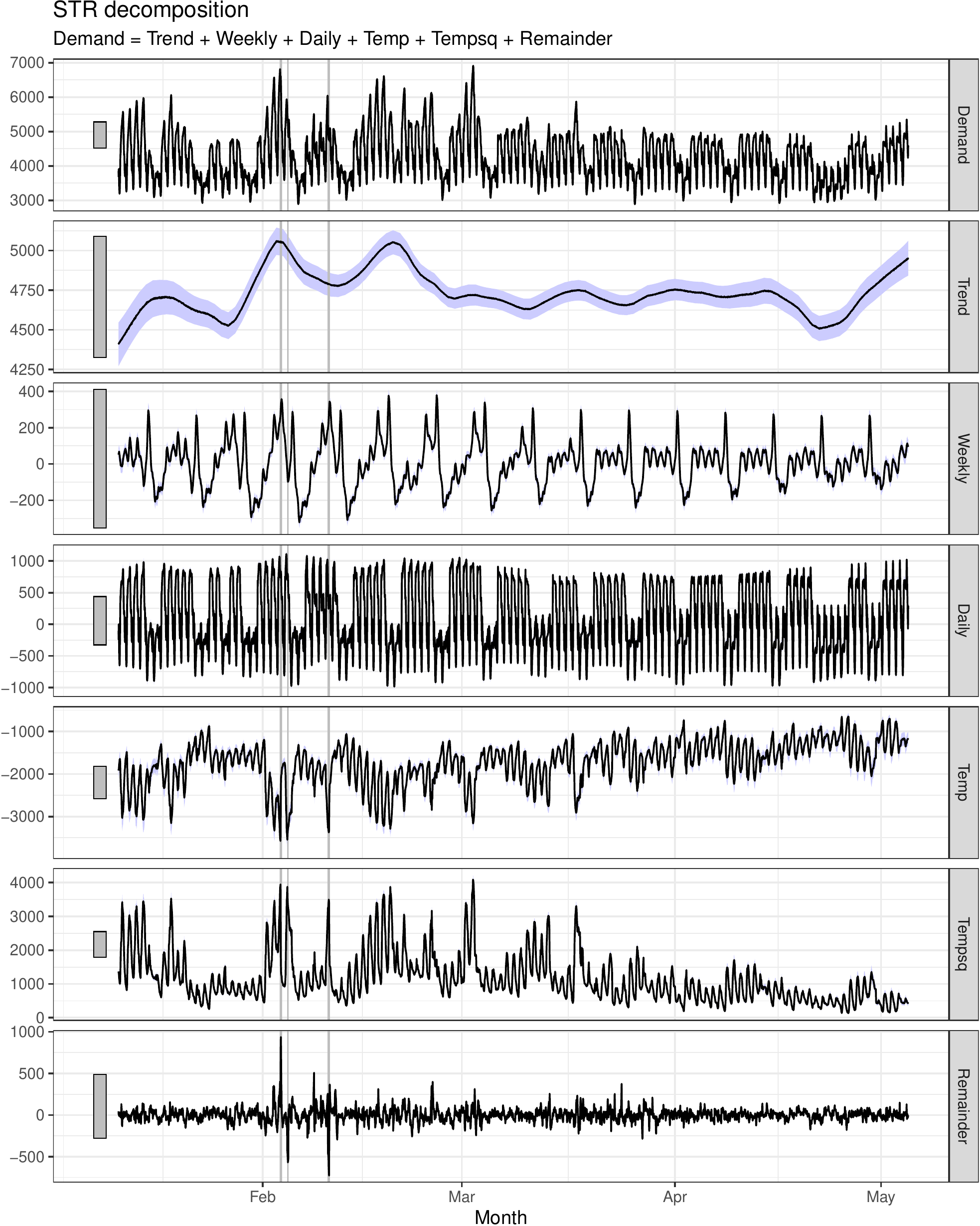}
\caption{\label{fig:vic}Electricity demand in Victoria, decomposed using STR. The original data are plotted in black in the first panel, with the decomposed trend plotted in red. The subsequent panels contain the weekly seasonal pattern; the daily seasonal pattern with complex topology; the effect of temperature in Melbourne, which has a daily seasonal pattern with complex topology; the effect of squared temperatures in Melbourne, which is allowed to vary over time; and the residuals. Grey vertical lines represent the residuals with the largest absolute values.}
\end{figure}

Figure \ref{fig:vic} shows the time series decomposed using STR.
The \(\lambda\) coefficients were chosen semi-automatically (the starting point for the minimization procedure was chosen according to experiments involving the minimization of the same problem with fewer predictors). Five-fold cross-validation with a gap length of \(g=336\) (one week) for the optimal smoothing parameters yielded the optimal mean squared error.

Two seasonal patterns and two regressors are used for the decomposition.
Thus, the data are represented as the sum of six components: trend, weekly seasonality, daily seasonality with a complex topology (work-day and non-work-day seasonality including transition periods), temperature and squared temperature (which are time-varying but non-seasonal), and the remainder. It has to be emphasized that the second seasonal component represents work-day, non-work-day seasonal patterns and the patterns of the transition periods between them as a single structure. To the best of our knowledge, no other seasonal decomposition method has this powerful ability. Another unique STR feature, illustrated in this decomposition, is the use of flexible regressors (components four and five). We could also have used a more complex approach with seasonal regressors (which is another unique feature of STR).

It is interesting to look carefully at the remainder term, as it will contain evidence of events that affect (or are correlated with) the electricity consumption, but which have not been captured in the model. The ten residuals that are largest in absolute value are shown in Table \ref{tab:outlierstable}, corresponding to the grey vertical lines in Figure \ref{fig:vic}.

\begin{table}[!bh]

\caption{\label{tab:outlierstable}The ten residuals that are largest in absolute value after an STR decomposition.}
\centering
\begin{tabular}[t]{rlcr}
\toprule
Date & Day of week & Time period & Residual\\
\midrule
3 February 2000 & Thursday & 14:30 -- 15:00 & 687.3\\
10
3 February 2000 & Thursday & 15:00 -- 15:30 & 871.3\\
10
3 February 2000 & Thursday & 15:30 -- 16:00 & 937.0\\
10
3 February 2000 & Thursday & 16:00 -- 16:30 & 767.2\\
10
3 February 2000 & Thursday & 16:30 -- 17:00 & 580.4\\
10
4 February 2000 & Friday & 16:00 -- 16:30 & -569.7\\
10
10 February 2000 & Thursday & 14:30 -- 15:00 & -611.4\\
10
10 February 2000 & Thursday & 15:00 -- 15:30 & -658.9\\
10
10 February 2000 & Thursday & 15:30 -- 16:00 & -695.7\\
10
10 February 2000 & Thursday & 16:00 -- 16:30 & -728.5\\
\bottomrule
\end{tabular}
\end{table}

\begin{figure}
\centering
\includegraphics[width=\textwidth]{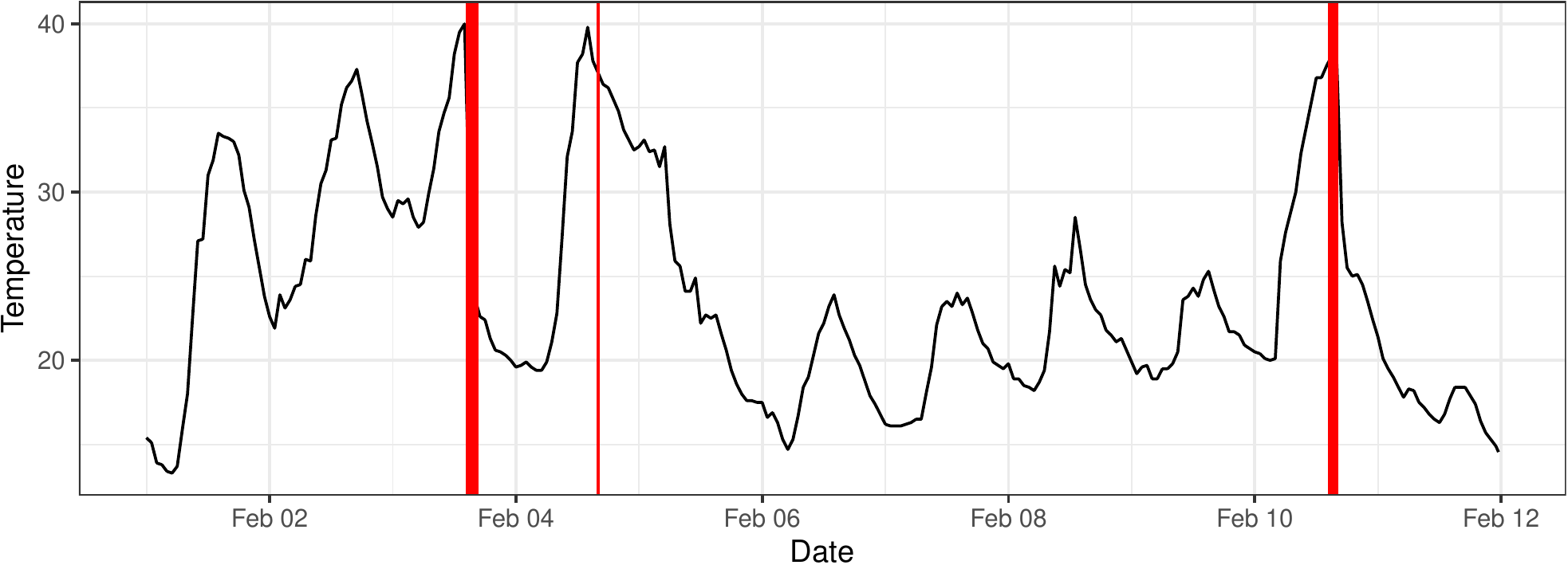}
\caption{\label{fig:MelTemp}The temperature in Melbourne on 1 February 2000, and the following ten days. Ten red lines (two groups of five lines each) mark the times of the outliers. The first group spans 14:30--17:00 on 3 February, while the second group spans 15:30--18:00 on the following day.}
\end{figure}

Melbourne is famous for its unstable weather, and the outliers can be explained by unusual weather on those days.
Five positive outliers occurred, starting from 2:30 pm on Thursday 3 February, 2000, probably because it was one of the hottest days in Melbourne (40\(^{\circ}\)C), but with a sudden drop in temperature (see the first five red lines in Figure \ref{fig:MelTemp}). We can speculate that although the temperature dropped at around 2:30 pm, the buildings stayed hot because they had heated up during the very hot previous three days and two nights. Thus, the electricity consumption was higher than that expected by the model, which only accounts for the contemporaneous temperature.

A negative outlier appeared at 4:00 pm on the following day, which was also very hot (39.8\(^{\circ}\)C) (see the separate single red line in Figure \ref{fig:MelTemp}). It is very unusual to have a hot day following a cool night in Melbourne. In this case, the buildings did not require as much air-conditioning as usual, due to the cooling effect of the previous night. Consequently, the model over-estimated the electricity consumption at that time.

The last four negative outliers started at 2:30 pm on Thursday 10 February, 2000, exactly one week later than the first four outliers. The explanation is similar to the previous case: it was a very hot day (38.1\(^{\circ}\)C) after a series of rather cool days and nights, and the model overestimated electricity consumption.

The model could be modified to better handle these situations by using lagged as well as concurrent temperatures, along with the change in temperatures over recent periods.

\hypertarget{extensions-and-discussion}{%
\section{Extensions and discussion}\label{extensions-and-discussion}}

Because we have used a linear modelling framework, it is very easy to extend our approach in many directions, exploiting the vast literature on linear regression models. We highlight a few possible extensions here.

\hypertarget{sec:func-coeff}{%
\subsection{Functional coefficients}\label{sec:func-coeff}}

We can reduce the dimension of the model by using a linear combination of smooth basis functions (such as splines) for the trend:
\begin{equation}
  \label{l_t}
  T_t = \sum_{j=1}^{q}\nu_j \psi_j(t),
\end{equation}
so that \(\bm\ell = \bm{\Psi}\bm{\nu}\), where \(\bm{\nu}\) contains coefficients to be estimated. Then no difference operators are needed to ensure smoothness. Instead, the value of \(q\) and shape of the basis functions \(\psi_j\) will determine the smoothness.

Similarly, seasonal functional components can be obtained, using Fourier terms for example \citep[reminiscent of][]{delivera2011forecasting}. This approach leads, with no additional effort, to handling seasonality with a fractional or varying period.

\hypertarget{robust-regularization}{%
\subsection{Robust regularization}\label{robust-regularization}}

If we replace the \(L_2\) norm used in \eqref{STR:ll}, with the \(L_1\) norm, we obtain a robust variant of STR. Equivalently, we replace the normality assumption with a double exponential distribution in the expression for the log-likelihood.

This can be written as a quantile regression \citep{koenker2005quantile}:
\begin{equation}
  \label{RSTR:qr}
  \hat{\bm\eta} = \operatorname{arg\,min}\limits_{\bm\eta} \left\|\bm{y}^{+}-\bm{X}\bm{\eta}\right\|_{L_1}\ ,
\end{equation}
where \(\bm{y}^{+}\), \(\bm{X}\) and \(\bm\eta\) are defined as in \eqref{STR:linear_model}. This can be solved numerically using a quantile regression algorithm \citep{quantregR}.

The \(L_1\) and \(L_2\) norms can also be mixed according to different assumptions regarding the distributions of the model components. In such cases, the minimization problem can be reduced to the LASSO minimization problem.

A mixture of norms can be useful, for example, in cases where the trend and seasonal patterns are smooth but the noise component has outliers. It can also be useful when the trend changes abruptly, but the seasonal components are smooth, and the noise is distributed normally (or at least has no outliers).

\hypertarget{correlated-residuals}{%
\subsection{Correlated residuals}\label{correlated-residuals}}

A variation of \eqref{STR:linear_model} allows the errors to be correlated with covariance matrix \(\bm{\Sigma}_\varepsilon\). Then the maximum likelihood estimates are given by the generalized least squares solution
\begin{equation}
\label{STR:solution_GLS}
\hat{\bm\eta} = (\bm{X}'\bm{\Sigma}_\varepsilon^{-1}\bm{X})^{-1}\bm{X}'\bm{\Sigma}_\varepsilon^{-1} \bm{y}^{+} ,
\end{equation}
with
\begin{equation}
\label{STR:beta_covar_STR}
\text{Cov}(\hat{\bm\eta}) = (\bm{X}'\bm{\Sigma}_\varepsilon^{-1}\bm{X})^{-1} .
\end{equation}

\hypertarget{level-shifts-and-interventions}{%
\subsection{Level shifts and interventions}\label{level-shifts-and-interventions}}

The STR approach allows us to deal easily with level shifts, spikes, and other intervention patterns as defined by \citet{BoxTiao1975}. Where these are linear terms, they can simply be added to the \(\bm{X}\) matrix in the usual way. For nonlinear intervention effects such as ``shift and decay'' patterns, non-linear least squares will need to be used.

\hypertarget{forecasting}{%
\subsection{Forecasting}\label{forecasting}}

The model can be used for forecasting by simply treating future observations as missing, and then estimating them. The smooth trend \(T_t\) will be continued linearly, much like a natural spline.

If the remainder term \(R_t\) is autocorrelated, the forecasts can be adjusted to allow for the short-term dynamics represented by the autocorrelations (e.g., by fitting an ARMA model to \(R_t\)). Prediction intervals can be obtained in the standard way for linear regression models.

When the STR model is being used for forecasting, better results will probably be obtained by selecting the smoothing parameters using \(K\)-fold cross-validation where the value of \(g\) should be comparable to the forecasting horizon \(h\) in order to avoid the trend being too flexible.

\hypertarget{feature-engineering}{%
\subsection{Feature engineering}\label{feature-engineering}}

The STR model provides a way to identify new features in time series data, such as weekly or daily patterns. These can then be used within other algorithms. For example, they could be used to classify time series \citep{catch22}, to find anomalous time series (as distinct from anomalous observations) \citep{cikm2015}, to select a forecasting model \citep{fforms}, or to create a weighted forecast combination \citep{fforma}.

\hypertarget{other-applications}{%
\subsection{Other applications}\label{other-applications}}

We have looked at two different business applications in monthly supermarket revenue and half-hourly electricity demand. It is easy to imagine many others, and we conclude with just three to illustrate the range of possibilities that could be used with this method.

\begin{enumerate}
\def\labelenumi{\arabic{enumi}.}
\tightlist
\item
  A retail outlet wishes to measure the effect of a promotion by studying the increased number of customers in their stores. The promotion took place during a week that includes a public holiday. They can use STR with seasonal, trend, and holiday effects, along with a categorical variable indicating the periods before, during and after the promotion. This gives a direct measure of the number of additional customers during the promotion, and the number of additional customers in the weeks following the promotion.
\item
  A web services company wants to identify potential hostile attacks by measuring unusual traffic on their servers. The usual traffic varies by time of day, day of week, time of year, whether there is a public holiday or not, and what the weather is like in the area. They can use STR with all of these seasonal variables and some weather variables, and look for unusually large values in the remainder series. These indicate traffic that is unusual after taking account of the regular causes of variation.
\item
  An energy company needs to forecast gas usage in a region for the next month, and they have ten years of hourly data. However, COVID-19 has caused changes in the usage profile as many people are working from home. STR can be used to estimate the trend, regular seasonal patterns with public holiday effects, and new seasonal patterns since people started working from home (including possibly new public holiday effects). The model can include temperature variables to allow for heating effects, and these can interact with the COVID-19 variables.
\end{enumerate}

\hypertarget{concluding-comments}{%
\subsection{Concluding comments}\label{concluding-comments}}

This article has introduced a new flexible approach to seasonal-trend decomposition using a linear regression model which can be used for a wide range of time series. The STR method allows for multiple seasonal periods and complex seasonality, missing values in input data, provides confidence intervals, finds smoothing parameters, and allows regressors to be taken into account with coefficients that may be time-varying and seasonal.

The main novelties of the STR approach are as follows. First, the seasonal components are represented as two dimensional structures of optionally complex topology. This is a completely new way of viewing seasonal components in the field of seasonal-trend decomposition. Second, the influence of covariates is allowed to be time-varying and can be seasonal, cyclic, etc. In all cases, the seasonal, cyclic or other components can have complex constraints which are difficult (if not impossible) to represent using other methods. Third, the approach proposes a regression perspective on time series decomposition, rather than the filtering or stochastic process approach which underpins most other methods. Fourth, beyond the methodological contributions mentioned above, the main theoretical contribution of the new approach is that it allows us to build a bridge between some stochastic processes and linear regression problems, which can give new mathematical interpretations and new computational methods for solving stochastic process problems. Finally, the new approach provides a unified framework for handling a very wide variety of seasonal-trend decomposition problems, which is easy to adapt in order to take into account further time series features.

An R package, \textbf{stR} \citep{stR}, implementing our model is available on CRAN.
The package also contains a vignette, which describes various practical aspects of the implementation in detail.

This paper was produced using Rmarkdown \citep{rmarkdown}. The source files to reproduce the paper, including all the examples, are available from \href{https://github.com/robjhyndman/STR_paper}{\texttt{github.com/robjhyndman/STR\_paper}}.

% Acknowledgments here
\ACKNOWLEDGMENT{Rob Hyndman gratefully acknowledges the support of the Australian Centre of Excellence for Mathematical and Statistical Frontiers.}

% References here (outcomment the appropriate case)

\bibliographystyle{informs2014} % outcomment this and next line in Case 1
\bibliography{strrefs} % if more than one, comma separated

% CASE 2: BiBTeX used to generate mypaper.bbl (to be further fine tuned)
%\input{mypaper.bbl} % outcomment this line in Case 2

%If you don't use BiBTex, you can manually itemize references as shown below.

\end{document}